\documentclass{ws-ijmpd}

\begin{document}

\markboth{N. Masetti et al.}
{The nature of unidentified {\it INTEGRAL} sources}

\catchline{}{}{}{}{}
%
%%%%%%%%%%%%%%%%%%%%%%%%%%%%%%%%%%%%%%%%%%%%%%%%%%%%%%%%%%%%%%%%%%%%

\vspace{-10cm}
\title{REVEALING THE NATURE OF NEW UNIDENTIFIED {\it INTEGRAL} SOURCES
%\footnote{For the title, try not to use more than 3 lines. Typeset the title in 10~pt Times roman, uppercase and boldface.}  
}

\author{NICOLA MASETTI, PIETRO PARISI\footnote{Also Dipartimento di 
Astronomia, Universit\`a di Bologna, Via Ranzani 1, I-40129 Bologna, Italy}, 
ELIANA PALAZZI, LOREDANA BASSANI, RAFFAELLA LANDI, ANGELA MALIZIA, 
FILOMENA SCHIAVONE, JOHN B. STEPHEN
%\footnote{Typeset names in 8~pt roman, uppercase. Use the footnote to indicate the present or permanent address of the author.}
}
\address{INAF -- Istituto di Astrofisica Spaziale e Fisica Cosmica di
Bologna, Via Gobetti 101, I-40129 Bologna, Italy\footnote{Address of 
corresponding author: masetti@iasfbo.inaf.it}}

\author{ELENA JIM\'ENEZ-BAIL\'ON}
\address{Instituto de Astronom\'{\i}a, Universidad Nacional Aut\'onoma de 
M\'exico, Apartado Postal 70-264, 04510 M\'exico D.F., M\'exico}

\author{VAHRAM CHAVUSHYAN}
\address{Instituto Nacional de Astrof\'{i}sica, \'Optica y Electr\'onica,
Apartado Postal 51-216, 72000 Puebla, M\'exico}

\author{LORENZO MORELLI}
\address{Dipartimento di Astronomia, Universit\`a di Padova, Vicolo 
dell'Osservatorio 3, I-35122 Padua, Italy}

\author{ELENA MASON}
\address{European Southern Observatory, Alonso de Cordova 3107, Vitacura,
Santiago, Chile}

\author{GASPAR GALAZ, DANTE MINNITI\footnote{Also Specola Vaticana, 
V-00120 Citt\`a del Vaticano}}
\address{Departamento de Astronom\'{i}a y Astrof\'{i}sica, Pontificia 
Universidad Cat\'olica de Chile, Casilla 306, Santiago 22, Chile}

\author{ANTONY J. BIRD, ANTHONY J. DEAN, VANESSA A. MCBRIDE}
\address{School of Physics \& Astronomy, University of Southampton, 
Southampton, Hampshire, SO171BJ, United Kingdom}

\author{PHIL A. CHARLES}
\address{South African Astronomical Observatory, P.O. Box 9, 
Observatory 7935, South Africa}

\author{ANGELA BAZZANO, PIETRO UBERTINI}
\address{INAF--Istituto di Astrofisica Spaziale e Fisica Cosmica di Roma, 
Via Fosso del Cavaliere 100, I-00133 Rome, Italy}

\maketitle

\begin{history}
\received{10 December 2009}
\revised{29 January 2010}
\comby{...}
\end{history}

\begin{abstract}
Since its launch on October 2002, the {\it INTEGRAL} satellite has 
revolutionized our knowledge of the hard X--ray sky thanks to its 
unprecedented imaging capabilities and source detection positional 
accuracy above 20 keV. Nevertheless, many of the newly-detected sources in 
the {\it INTEGRAL} sky surveys are of unknown nature. However, the 
combined use of available information at longer wavelengths (mainly soft 
X--rays and radio) and of optical spectroscopy on the putative 
counterparts of these new hard X--ray objects allows pinpointing their 
exact nature. Continuing our long-standing program running since 2004
(and with which we identified more than 100 {\it INTEGRAL} objects)
here we report the classification, through optical spectroscopy, of 25 
unidentified high-energy sources mostly belonging to the recently 
published 4$^{\rm th}$ IBIS survey.
\end{abstract}


\section{Introduction}

One main objective of the {\it INTEGRAL} hard X--ray satellite is the 
regular survey of the whole sky at high energies (above 20 keV). This 
makes use of the unique imaging capabilities of the IBIS 
instrument\cite{ibis} which permits the detection of sources at the mCrab 
level with a typical localization accuracy of 2-3 arcmin above 20 keV. A 
substantial fraction of the remaining objects ($\sim$30\%; see e.g. 
Ref.~\refcite{ajb}) had no obvious counterpart at other wavelengths and 
therefore cannot be associated with any known class of high-energy 
emitting sources.

To this aim, since 2004 our group has been actively performing a 
successful observational campaign for the optical identification of 
these unidentified objects: up to now we identified more than 100
{\it INTEGRAL} sources and found that about half of them are Active
Galactic Nuclei (AGNs), mostly located in the nearby Universe (at 
redshift $z \lesssim$ 0.1; see Ref.~\refcite{maso} and references 
therein).

Here we present the continuation of this work with the use of the
new {\it INTEGRAL} detections of unidentified sources reported
in the recently published 4$^{\rm th}$ IBIS survey.\cite{ajb}

\section{Sample selection}

Using the criterion applied to our past works (see e.g. 
Ref.~\refcite{maso}), we positionally cross-correlated the 208 
unidentified objects belonging to the 4$^{\rm th}$ IBIS survey\cite{ajb} 
with X--ray ({\it ROSAT}, {\it XMM-Newton}), radio (NVSS, SUMSS, MGPS) and 
far-infrared (IRAS) catalogues -- available online using 
SIMBAD\footnote{http://simbad.u-strasbg.fr} -- and with archival X--ray 
observations ({\it Swift}, {\it Chandra}). This was made in order to 
reduce the source error circle and pinpoint the putative optical 
counterpart. A few cases in which a catalogued emission-line optical 
object is present in the IBIS error box were also considered. We also 
included in our sample the source IGR J01054$-$7253, not included in the 
4$^{\rm th}$ IBIS survey but recently discovered during an {\it INTEGRAL} 
key programme observation,\cite{boz} and for which an arcsec-sized X--ray 
position obtained with {\it Swift} is available in the 
literature.\cite{coe}

This allowed us to pick out, and perform optical spectroscopy on, 25 
candidate counterparts. The list of selected objects is reported in the 
1st column of Table 1.

\section{Caveats}

While it is known\cite{jbs} that the presence of a single, bright soft 
X--ray object within the IBIS error circle indicates that it is with a 
high probability the lower-energy counterpart of the corresponding {\it 
INTEGRAL} source, the same cannot be said for radio and far-infrared 
sources, or for optically peculiar objects. Thus, for the IBIS sources 
which were selected using catalogues at wavelengths longer than soft 
X--rays, we caution the reader that the association with the selected 
optical object should await confirmation via a pointed observation with a 
soft X-ray satellite such as {\it XMM-Newtom}, {\it Chandra} or {\it 
Swift}. To stress this issue, when needed, we put an asterisk beside the 
source name in Table 1.

\section{Observations}

In this work, 6 telescopes located in Northern and Southern Hemisphere 
observatories were used; moreover, spectra from the 6dF\cite{6df} and 
SDSS\cite{sdss} on-line archives were also employed. In the following, the 
list of the telescopes used for the classification reported in Table 1 is 
given:

\begin{itemlist}
\item 1.5m telescope at CTIO, Chile;
\item 1.5m "Cassini" telescope in Loiano, Italy;
\item 1.8m "Copernico" telescope in Asiago, Italy;
\item 1.9m "Radcliffe" telescope at SAAO, South Africa;
\item 2.1m telescope in San Pedro M\'artir, Mexico;
\item 2.5m telescope at Apache Point Observatory, New Mexico, USA, for 
the SDSS spectra;
\item 3.58m NTT telescope at ESO-La Silla, Chile;
\item 3.9m AAT telescope of the Anglo-Australian Observatory in Siding Spring, 
Australia, for the 6dF spectra.
\end{itemlist}

\section{Results}

Table 1 reports, for each selected object, the corresponding 
classification obtained through optical spectroscopy, the object redshift 
and the name of the telescope used for the identification; Fig. 1 reports 
the spectra of the optical counterparts of some of these 25 selected IBIS 
sources.

\begin{table}%[ph]
\tbl{The sample of 25 {\it INTEGRAL} objects identified in this work 
together with their classification, redshift and telescope with which the 
identification was obtained. The asterisks indicate sources for which an 
X--ray position is still not available (see Sect.~3).}
{\begin{tabular}{@{}llll@{}} \toprule
Object name & Class & redshift & Telescope \\ \colrule
IGR J00158+5605         & Sy1.5      & 0.168 & SPM    \\
IGR J00465$-$4005       & Sy2        & 0.201 & CTIO   \\
IGR J01054$-$7253       & HMXB       & 0     & SAAO   \\
IGR J01545+6437$^*$     & Sy2        & 0.034 & Asiago \\
IGR J02086$-$1742       & Sy1.2      & 0.129 & SPM    \\
IGR J05253+6447         & likely Sy2 & 0.071 & Loiano \\
1RXS J080114.6$-$462324 & CV         & 0     & NTT    \\
MCG +04$-$26$-$006      & LINER      & 0.020 & SPM    \\
IGR J1248.2$-$5828      & Sy1.9      & 0.028 & SAAO   \\
IGR J13187+0322$^*$     & QSO        & 0.606 & SDSS   \\
IGR J14301$-$4156$^*$   & Sy2/LINER  & 0.039 & 6dF    \\
IGR J15311$-$3737       & Sy1        & 0.127 & SAAO   \\
IGR J15549$-$3739$^*$   & Sy2        & 0.019 & 6dF    \\
IGR J16287$-$5021       & LMXB       & 0     & NTT    \\
IGR J16327$-$4940$^*$   & HMXB       & 0     & SAAO   \\
1RXS J165443.5$-$191620 & CV         & 0     & SPM    \\
IGR J18311$-$3337$^*$   & Sy2        & 0.066 & 6dF    \\
IGR J19077$-$3925       & Sy1.9      & 0.073 & 6dF    \\
IGR J19113+1533$^*$     & HMXB       & 0     & SPM    \\
IGR J19118$-$1707$^*$   & Sy2/LINER  & 0.024 & 6dF    \\
PKS 1916$-$300          & Sy1.5/1.8  & 0.167 & 6dF    \\
IGR J19552+0044         & CV         & 0     & Loiano \\
1RXS J211336.1+542226   & CV         & 0     & SPM    \\
1RXS J211928.4+333259   & Sy1.5/1.8  & 0.051 & SPM    \\
1RXS J213944.3+595016   & Sy1.5      & 0.114 & SPM    \\ \botrule
\end{tabular}}
\end{table}

\begin{figure}%[pb]
\vspace{-2cm}
\centerline{\psfig{file=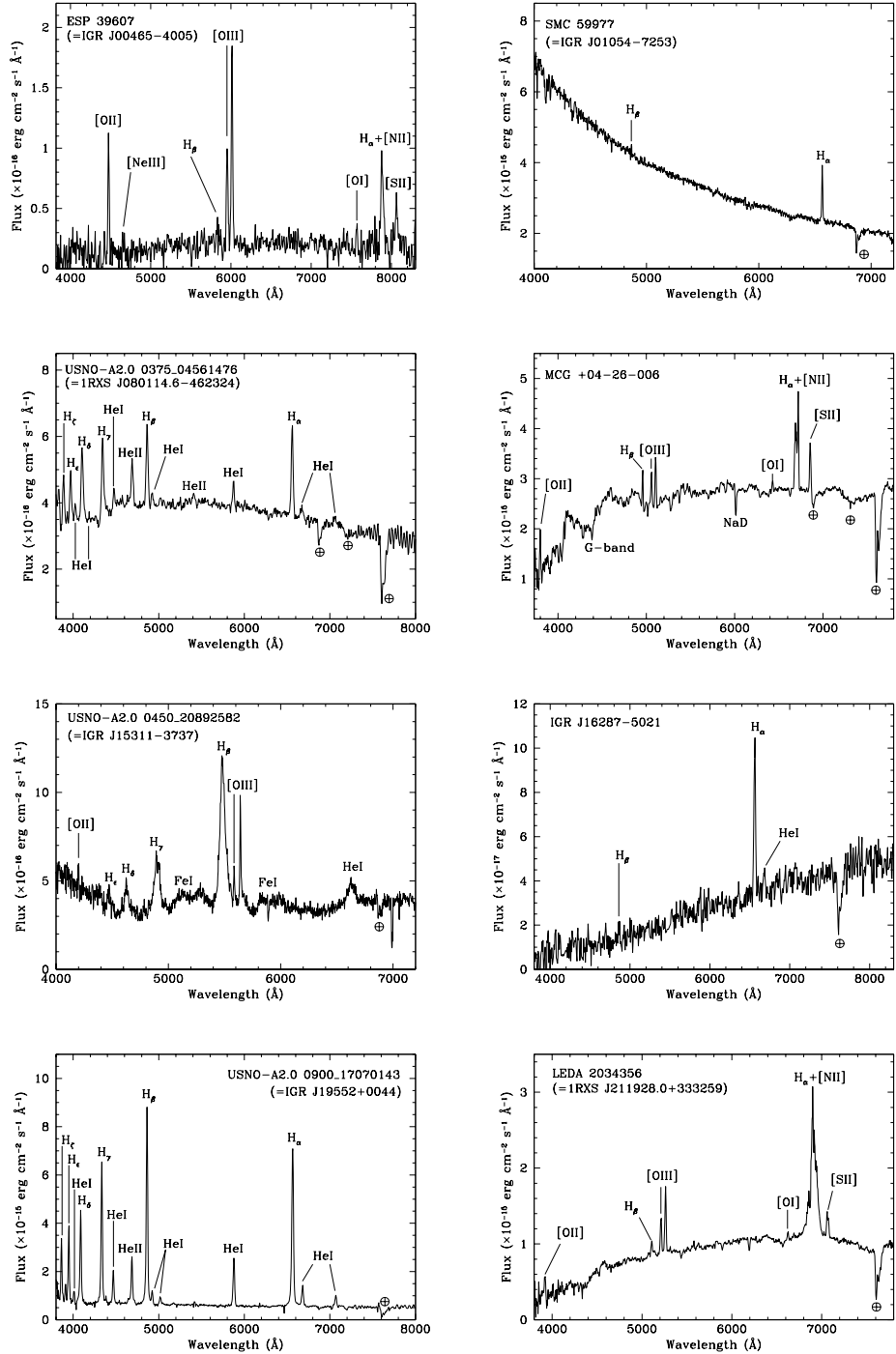,width=17cm}}
%\vspace*{8pt}
\vspace{-3.5cm}
\caption{Optical spectra of the counterparts of 8 objects belonging 
to the sample of 25 {\it INTEGRAL} sources identified in this work and 
listed in Table 1. These data allowed us to securely determine the nature 
and the redshift of these objects through inspection of their absorption 
and emission spectral features. The spectra are not corrected for the 
intervening Galactic absorption. For each spectrum the main spectral 
features are labeled. The symbol $\oplus$ indicates atmospheric telluric 
absorption bands.}
\end{figure}

As one can see from Table 1, most objects (17, i.e., 68\%) have an
extragalactic origin, and are nearly equally divided into Type 1 and Type 
2 AGNs (9 and 8 cases, respectively). Six of them (5 of Type 1 and only 1 
of Type 2) have $z >$~0.1, indicating that this new, deeper IBIS survey is 
able to detect more distant AGNs.  Only 32\% (8 sources) of our 
identifications are instead of Galactic nature. Interestingly, half of 
them are (likely magnetic) Cataclysmic Variables (CVs): these 
identifications increase by 13\% the number of hard X-ray emitting CVs 
detected with {\it INTEGRAL} up to now (see Ref.~\refcite{ss}).

\section{Conclusions}

We here presented the identification of a first set of 25 sources of 
unknown or uncertain nature, 24 of which belonging to the 4$^{\rm th}$ 
IBIS survey. We found that 2/3 of them have an extragalactic nature, with 
redshifts in the range 0.019--0.606, while the remaining ones are Galactic 
sources. It is noteworthy that the majority of the latter ones are 
(possibly magnetic) CVs.

These preliminary results further confirm the {\it INTEGRAL} capabilities 
of detecting AGNs and hard X-ray emitting CVs. We also stress that with 
this work we already reduced by 12\% the number of unidentified sources of 
the 4$^{\rm th}$ IBIS survey, bringing it from 208 to 184.

\section*{Acknowledgments}

This research has made use of the SIMBAD database operated at CDS, 
Strasbourg, France, of the NASA/GSFC's HEASARC archive, and of the 
HyperLeda catalogue operated at the Observatoire de Lyon, France.
We thank the referee for useful indications.
The authors acknowledge the ASI and INAF financial support via grant
No. I/008/07. PP is supported by the ASI-INTEGRAL grant No. I/008/07.
VC is supported by the CONACYT research grant 54480-F (Mexico).

%\begin{thebibliography}{000} %for 3 digits
%\begin{thebibliography}{00}  %for 2 digits

\end{document}